\documentclass[letterpaper, 10 pt, conference]{ieeeconf}  

\IEEEoverridecommandlockouts                              

\usepackage{graphicx} 
\usepackage{cite}
\usepackage{paralist}
\usepackage{hyperref}
\hypersetup{breaklinks=true,hidelinks}
\usepackage{algorithm}
\usepackage{algpseudocode}
\usepackage[table,xcdraw]{xcolor}

\title{\LARGE \bf
Anxolotl, an Anxiety Companion App - Stress Detection
}

\author{Nuno Gomes$^{1}$ \and Matilde Pato$^{2}$ \and Pedro Santos$^{1}$  \and André Lourenço$^{3}$  \and Lourenço Rodrigues$^{3}$ 
\thanks{*This work was not supported by any organization}
\thanks{$^{1}$Nuno Gomes and Pedro Santos are with ISEL - Instituto Superior de Engenharia de Lisboa, Instituto Politécnico de Lisboa, Lisboa, Portugal
        {\tt\small gomes.nunoms@gmail.com, pedro.santos@deetc.isel.ipl.pt}}%
\thanks{$^{2}$Matilde Pato is with FIT-ISEL - Instituto Superior de Engenharia de Lisboa, Instituto Politécnico de Lisboa and LASIGE, FCUL, Universidade de Lisboa, Lisboa, Portugal
        {\tt\small matilde.pato@isel.pt}}%
 \thanks{$^{4}$André Lourenço and Lourenço Rodrigues are with ISEL - Instituto Superior de Engenharia de Lisboa, Instituto Politécnico de Lisboa and CardioID technologies, Lisboa, Portugal
    {\tt\small andre.lourenco@isel.pt, lourenco.rodrigues@isel.pt}}%
}

\begin{document}

\maketitle
\thispagestyle{empty}
\pagestyle{empty}

\begin{abstract}
    Stress has a great effect on people's lives that can not be understated. While it can be good, since it helps humans to adapt to new and different situations, it can also be harmful when not dealt with properly, leading to chronic stress. The objective of this paper is developing a stress monitoring solution, that can be used in real life, while being able to tackle this challenge in a positive way. The SMILE data set was provided to team Anxolotl, and all it was needed was to develop a robust model. We developed a supervised learning model for classification in Python,
    presenting the final results of $64.1\%$ in accuracy and a f1-score of $54.96\%$. The resulting solution stood the robustness test, presenting low variation between runs, which was a major point for it's possible integration in the Anxolotl app in the future. The code is available at \url{https://github.com/matpato/CfP-Workshop-and-Challenge-Wellbeing.git}.        

\end{abstract}


\section{Introduction}\label{sec:introduction}

Stress is a term that describes bodily reactions to perceived physical or psychological threats \cite{jongyoon2012development}. Since the start of stress level recording among the population, these values have been on the rise, and the pandemic had a significant impact on them. There is a consistent increase of stress-related mental symptoms (anxiety, depression, general psychological distress) in the general population during the pandemic compared to before \cite{eu2021mentalhealth}. 

While these facts are dire, stress in its inception is a good evolutionary response to dangerous situations, allowing our bodies to be better prepared to perform in the face of a ``fight-or-flight'' situation. An example of such a situation could be an encounter with a tiger. Nowadays, it is unusual to find tigers in a person's day-to-day life, and so, it is more prevalent in the case of deadlines or responsibilities, and its purpose is to help humans to be better prepared to deal with such events, using biological changes to face a recognized threat. It still can be beneficial, keeping us alert in dangerous situations and focused to meet challenges \cite{jongyoon2012development}.

On the other side, if such situations keep adding up and stress does not subside, it stops being classified as an acute stress response, and it starts entering the chronic stress realm. At this stage, our bodies are producing hormones to keep the stress response up, but the outcome starts being more negative than positive. This chronic stress can lead to the atrophy of the brain mass and decrease of its weight. These structural changes bring about differences in the response to stress, cognition and memory\cite{habib2017stressimpact}

Mental health problems exist along a continuum, from mild, time limited stress, to severe mental health conditions, and while mental illnesses and stress are not the same, they are closely related. Stress and anxiety affect most people at some point in their lives, but the regularity at which that happens is one of the key points of classifying such events as a disease. Focusing on the anxiety and anxiety disorders, they are the most common type of mental illness in the world, affecting 264 million people worldwide as of 2017, with an increase of $14.9\%$ per decade \cite{elgendi2019assessinganxiety}. While, the rise of both stress and anxiety is related, so are their symptoms.

Anxiety is one of the most pervasive and ubiquitous human emotions, in all cultures \cite{sarason1990testanxiety}. It is considered a basic negative emotion, such as sadness, anger, worry and fear. Anxiety, fear and stress all share similarities and might even overlap to some extent, but they are different states:  Stress has a clear cause, which is called a stress-causing factor or a stressor, such as the tiger mentioned before. Fear also shares some similarities to stress, but it is classified as an emotion and might trigger a stress response, it is associated with danger and/or insecurity, and it is also focused on immediate present danger. Anxiety, by contrast, corresponds to a state of uncertainty, and it is more closely
related to a future-oriented mood state associated with preparation for possible, up-coming negative event.

Measuring anxiety and stress has a big overlap,  due to a propensity of one to cause the other, common risk factors, as well as the bodily reactions being similar. Choosing which biological data to capture and analyze to target each situation becomes paramount for detection. Nevertheless, the reason for the association between these psychological syndromes is yet to be established \cite{ramon2020prevalencedepression}. Regarding their monitorization, context is likely to be utterly important, since it allows for a better evaluation of the data, and questionnaires can fill the gap in distinguishing both, as presented in \cite{bickman2020improvingmental}.

The symptoms of these conditions can be divided into Somatic (physical) and Psychic. For the most part, the symptoms most commonly associated with each are:
\begin{itemize}
    \item Anxiety
    \begin{itemize}
        \item Somatic --- tremors, palpitations (increased or irregular heart rate), dizziness, nausea, shortness of breath, sweating, muscle tension, etc.;
        \item Psychic --- difficulty concentrating, nervousness, Insomnia, constant worry, etc.;
    \end{itemize}
    \item Stress
    \begin{itemize}
        \item Somatic --- aches and pains, palpitations, muscle tension, digestive problems, etc.,
        \item Psychic --- anxiety, irritability, depression, sadness, panic attacks, etc..
    \end{itemize}
\end{itemize}

Looking at the symptoms, we can see some overlap. Given that stress can cause an anxiety response, then all the symptoms present in anxiety become targetable on stress detection. Current consumer wearables are not yet capable of distinguishing data with such precision, yet they are the most accessible way of monitoring both cases.

The production of smart devices to help individuals monitor components of their health has been on the rise during the last few years \cite{hickey2021smartdevices}. The presence of smartphones among the population is almost universal, and these two tools could be used as a way of bringing comfort and quality of life to people suffering from mental illnesses such as anxiety or chronic stress.

\subsection*{Contributions}

The Anxolotl project focuses on trying to supplement a more nuanced solution to a very nuanced problem, which is the management of mental health and follow-up of mental illnesses, namely General Anxiety Disorder (GAD) and Panic Disorder (PD). By taking advantage of consumer grade wearables, which are already very present worldwide, and using them to allow patients to better manage their mental health and well-being. The big focus is to provide a support tool, one mainly used to keep track of their mental health data, and allow them to intervene before an acute crisis settles, or chronic state in the case of stress.

\subsection*{Anxolotl - An Anxiety Companion App}

The Anxolotl --- An anxiety companion app ---  presents a system that can reliably detect anxiety and stress levels, detect panic attacks (PAs) as long as a wearable is being used. Ideally, upon the detection of abnormal anxiety or stress levels, a notification would pop up, and in the case of detection of PAs, the user will be able to choose which mechanisms to use, such as automatically calling a selected person, buzzing or suggesting breathing exercises. The app is intended to run on the background and auto-start to be as frictionless as possible to use. 

The main idea is to give users control over their mental health situation. This would translate into being able to check anxiety and stress levels on a smartphone, as well as being warned by a notification in the case of consistently high stress or anxiety levels. Short term solutions such as meditation, or wellness exercises could be suggested, but the main point is the detection. As long as the user leaves the app on the background, and wears the wearable, these mental health statistics can be recorded, and the user can live its life ignoring the app, until the time the app detects an abnormality.

Finally, there are some non nuclear objectives, such as the presentation of the data to a validated medical professional, and stress detection. The last one is discussed in this paper, as well as the algorithm used. It is intended to work along the anxiety detection in the way of giving users control over their situation. As stated before, anxiety and stress share symptoms, and to address this issue, questionnaires will be used such as the GAD questionnaire (GAD-7), and for stress the Perceived Stress Questionnaire (PSS). On this paper, we focus on the stress detection without context, which is tougher, given the lack of truly unique symptoms.

\subsubsection*{Environment}

The Anxolotl project starts by capturing data from the wearable. That data is sent in real time to a smartphone via Bluetooth Low Energy. The app is developed in Flutter to allow interoperability between iOS and Android, having a wider reach.

\begin{figure}[t]
    \centering
    \framebox{\parbox{3in}{\includegraphics[width=3in]{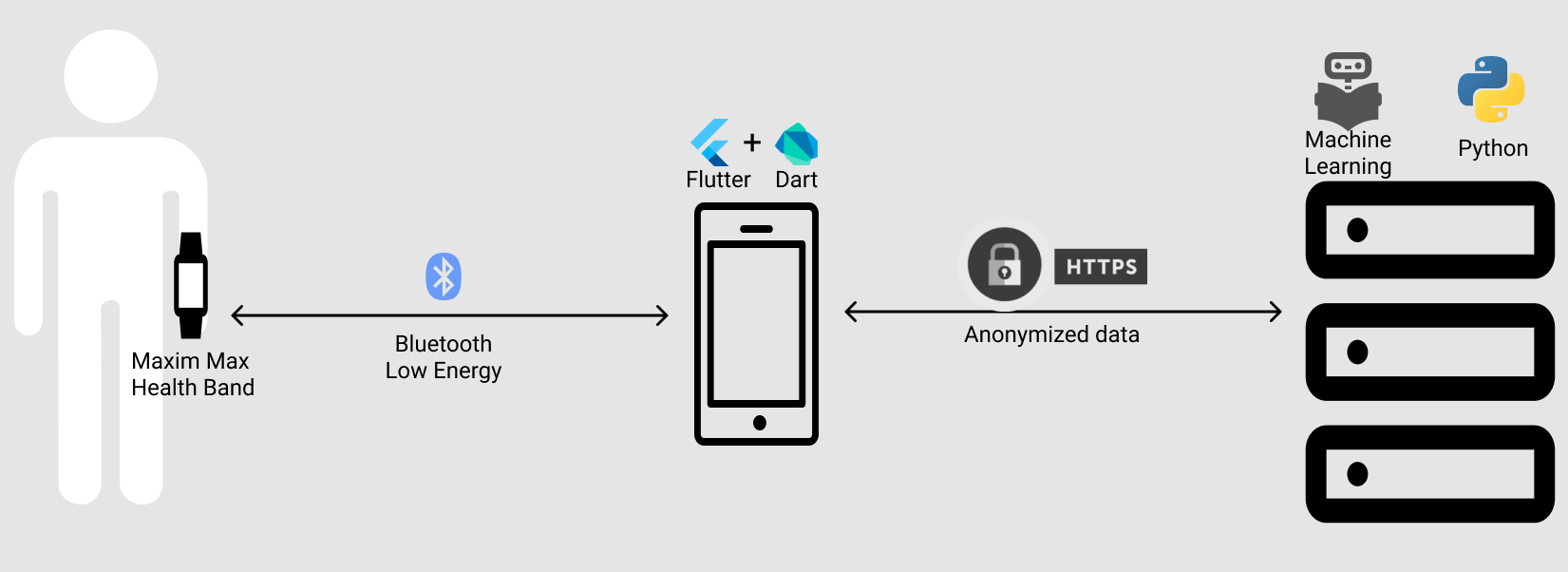}}}
    
    \caption{Anxolotl solution designed environment}
    \label{fig:anxolotlSystem}
 \end{figure}

As shown in the Fig. \ref{fig:anxolotlSystem}, filtering is applied (low-pass) on the mobile app,  removing any erroneous data, and some data processing is done as well. Then, the data is synced every $10$ min to the data center via HTTPS. The datacenter contains the models on an initial stage, where they are trained. On a later iteration these trained models could be loaded and applied on the smartphone to reduce latency and have a real time response. The data center is responsible of receiving the biological data and training classifier models with it, giving the mental health statistics in return. A response is then sent to the smartphone identifying stress levels and presenting them to the user. The user can then check their mental health levels on their smartphone, as well as receive notifications when the models detect unusually high levels.

\subsection*{Organization}
The remainder of this paper is structured as follows: in the section \ref{sec:relatedwork} we give an overview of the current existing research on the stress subject, as well as mention some relevant scientific projects. Section~\ref{sec:background} describes the technical context needed for an easier understanding of our solution. Section~\ref{sec:methodology} describes the methodology we use for the model design, as well as iterate on the different options. Section \ref{sec:results} presents the results and discusses some of the limitations associated with our work. Finally, Section \ref{sec:conclusion} wraps up all the work, and presents our findings as well as the next steps.

\section{Related Work}\label{sec:relatedwork}

Lately, there has been a push towards a better mental health maintenance, since it is detrimental to an individual's quality of life. This section will focus on the papers that, as of late, provided good results with wearable compatible sensors in measuring stress, anxiety or panic attacks. Given the focus of this paper, a good collection of sensors data to measure each of this metrics will bring immense value, since it will allow for more combinations of sensors to be picked. The pioneers in this field were Healey and Picard who showed, in 2005, that stress could be detected using physiological sensors \cite{gjoreski2017monitoringstress}. 

The purpose of Healey and Picard \cite{healey2005detectingstress}
was  to distinguish between $3$ base levels (\textit{low}, \textit{medium} and \textit{high}) of stress in drivers, with an accuracy rate of around $97\%$. The stress addressed here was the stress with a negative bias, namely distress. Four types of physiological sensors were used during the experiment: 
\begin{inparaenum}[]
\item electrocardiogram (ECG), 
\item electromyography (EMG), 
\item skin conductivity (also known as SC, electrodermal activation (EDA) or galvanic skin response (GSR)), and 
\item respiration (through chest cavity expansion). 
\end{inparaenum}
Their algorithm included the mean and variance of the EMG 
taken in the hand, respiration and the mean of the heart rate (HR) over one second intervals throughout a drive. In this paper, the best correlating signals with stress levels were the mean of the skin conductivity ($.47$), followed by the L100 (frequency domain HR Variability (HRV)) ($.41$) and finally the HR ($.30$). According to this paper, using the HR and GSR with intervals of $5$ min., stress levels accurately could accurately be predicted $97.4\%$ of the time.

Hee Han et al. \cite{Han2020objectivestress} focused on measuring the stress levels from a population of $17$ subjects on an everyday setting and on a laboratory setting, while binarily accessing their stress condition (stressed or not-stressed). This paper focused on using $3$ sensors, photoplethysmography (PPG), ECG and GSR. In a lab setting, the paper provided a $94.6\%$ accuracy in distinguishing the stress levels, while that figure dropped to $81.82\%$ on an everyday setting. One of the outcomes was finding that in the everyday setting, GSR + ECG group showed the best everyday accuracy, which was $90.91\%$ \cite{Han2020objectivestress}. Another finding was that the sensors from the wearables tend to perform worse on an everyday setting, since data capture noise became a real problem when it came to measuring data in an ordinary setting.

Finally, an overlook of the current situation in measuring stress, we can see value in all the presented sensors. While HRV and PPG are relatively recent, they are also promising, as HRV was identified as the most useful physiological metric for detection of stress and anxiety \cite{hickey2021smartdevices}, it was also observed that HR and GSR were the most regularly used sensory signals because they gave the most promising results and high-accuracy for detecting stress and its
levels\cite{gedam2021reviewmental}.

\section{Background}\label{sec:background}

Here we present the context we think is necessary to understand both parts of this work, both the more medical, as well as the more technological. Stress can be measured by monitoring physiological indicators such as heart activity, blood activity, skin response (GSR) or skin temperature (ST), and we address this problem  with a strong theoretical background. While measuring stress on itself is tricky, we can measure indicators of such, and such events must be explained and theoretically correlated with stress itself. Regarding the machine learning (ML), we will also present the algorithms and methods we consider important. While the development is highly empirical, given a ML context, it highly relies on a basic understanding of the human body, and the relations between stress and stress related physiological data.

\subsection*{Measuring biological data}

Given that stress is a bodily response to a \textit{stimulus}, or multiple \textit{stimuli} with somatic symptoms, those same symptoms can be measured. Multiple types of symptoms allow the existence of multiple different ways of measuring, and while the most promising data seems to be heart activity and galvanic activity related \cite{gjoreski2017monitoringstress,memar2021stressclassification,Han2020objectivestress} there is also value in the monitoring of ST. All these factors play a role in the physical manifestation of stress on the human body, and these studies presented good results with accuracies of more than $90\%$ using the presented physiological data.

GSR refers to electrical changes that arise when the skin receives specific signals from the brain. These changes may be due to emotional activation, cognitive workload or physical exertion\cite{gjoreski2017monitoringstress}. While these changes can be subtle, stress can also cause sweat to happen, and as such, increase the level of GSR, which can be used for detection.

Heart activity is the most known of these biological signals, and most wearable devices can capture HR and HRV. While the HR increases upon stress, it also increases on many other ordinary phenomena, such as a scare, on the other hand HRV has a tighter relation with stress. Usually HRV is extrapolated from PPG and highly related to HR, and time-domain indices of HRV can quantify the amount of variability in measurements of the period between successive heartbeats, the Inter-Beat-Interval (IBI)\cite{gedam2021reviewmental}.

The "fight-or-flight" response restricts the blood flow from the extremities and increases the blood flow to vital organs. This peripheral vasoconstriction produces changes in ST on the extremities including hands, which can indicate stress and its intensity\cite{gjoreski2017monitoringstress}. While rises and drops in temperature are normal body functions, when correlated with other signals, ST might be a good indicator of a stress response, by using mean temperature or the slope of the temperature during a certain time frame.

It was observed that when stress occurs, HR, blood pressure, respiration rate, and GSR tend to increase while HRV and ST decrease \cite{gedam2021reviewmental}. This is not much different from a physical exercise session, and here is where context can make or break a model. But with this in mind, a good amount of features will bring better results to a model, given not a single feature can accurately detect stress.

\subsection*{Machine Learning}

In a complex problem such as stress detection, the application of some type of machine learning (ML) algorithms makes sense. The vast amount of data in a context were multiple variables, such as HR, HRV, GSR and ST, might have different outputs based on each other, makes it a prime target for the ML approach. It is no wonder it has already been applied to it, and continues to be used and researched to this day.

\subsubsection*{Feature Selection}
Machine learning algorithms are built with data that is fed to them, so it is easy to assume that the quality of the models is proportional to the amount and quality of information that is consumed. To take out any irrelevant information, it is common to apply a pre-processing step known as feature selection, in order to improve the model's performance. Following, are some of the techniques used. 

Recursive feature elimination (RFE) with cross-validation is one of the algorithms used to achieve this feature selection. Recursion is the process of repeating a process multiple times. In the case of \textit{RFE}, the process consists in generating a different model, and for each iteration different features are taken away based on the generated metrics. While this process takes place, the impact of the removal of each feature is observed in the model’s accuracy, to find the optimum set of features to use for the maximum results. 

Sequential Feature Selector adds (forward selection) or removes (backward selection) features to form a feature subset in a greedy fashion. At each step, this estimator chooses the best feature to add or remove based on the cross-validation score of an estimator. 

\subsubsection*{Classifiers}
Classification consists in predicting the class of a set of given data points; classes are sometimes called targets/labels or categories. Classification is the task of approximating a mapping function from input variables (X) to discrete output variables (Y). There are a lot of classification algorithms available, however, what dictates whether they perform accurately or not depends on the nature of the given data set and the relationships between data. Some of the most common classification algorithms are Support Vector Machines (SVM), K-Nearest Neighbours (K-NN), Random Forests (RF), Decision Trees and Naive Bayes (NB).

\subsubsection*{Classifiers in stress detection}

Many studies have applied multiple methods in feature selection and classification, but no universal algorithm has been developed for stress. With that in mind, it is a good idea to look at what came before to have a clear perspective in where to start regarding this subject.

In Gjoreski et al. \cite{gjoreski2017monitoringstress}, the best result for context stress detection, regarding F1-Score was the Decision Tree with $90\%$, followed by Random Forest with $74\%$, SVM with $69\%$ and K-NN with the same result. All these results were made in an aggregation-window with $10$ seconds. The no-context events had lower precision scores, around $7\%$ for true positives in the best model.

Memar and Mokaribolhassan \cite{memar2021stressclassification} presents a table, with a stress analysis review. Here, for data sets without context and using the data available in
our dataset (HR, GSR and ST), the best results in terms of accuracy were from a SVM with $80\%$, K-NN
with $88.6\%$ and Logistic Regression with $91.4\%$. On the other side, the measurements had a big number of sensors, which we do not have.

Lastly, Han et al. \cite{Han2020objectivestress}, had success using PPG, ECG and GSR, while classifying the stress with K-NN (multiple variables) and SVM, with accuracies ranging from $85\%$ to $95\%$, coming closer to $85\%$ on contextless stress detection on day to day tasks. Feature selection was used to reach those results, and classifier were tested along the development as well.

\section{Methodology}\label{sec:methodology}
The approach we use in this challenge is heavily influenced by current literature. Instead of a traditional heuristic approach based on a machine learning (ML) problem, this challenge is interpreted as a data problem. Since the provided data is not unprocessed, we opt to interpret the problem this way to try and connect the data we already have with the results we are aiming for.

While the influence of the literature is going to be relevant, another relevant feature of our work is the usage of the SKlearn framework, which brings some limitations, namely not having implementations of the most technically advanced algorithms, such as deep neural networks. With that said, we have to forfeit some of the more advanced algorithms, and focus on long established algorithms.

This section regards our analysis of the data set, features and their selection as well as an introduction to our ML algorithm choices.

\subsection*{Data}

Our solution uses the SMILE data set \cite{smileDataset}, and by extension it is designed to work well with it. A total of $45$ healthy adult Belgian participants ($39$ females and $6$ males) were recruited for SMILE. Among participants, the average age was $24.5$ years old and the standard deviation was $3.0$ years. On average, each participant contributed $8.7$ days of data. 
Two types of sensors were used for the data set, one for HR, and another one for GSR and ST. 

Regarding the data set itself, the data is not the original recorded data. It is anonymous and was reconstructed from a model based on the original data, and this process was made for the continuous portion of data by the data set providers. For the handcrafted portion of the data, we have $60$ min. of measurements \textit{per} stress label, which means the data has to be processed to fit into a 1-1 model --- one data point to one label. 

The data also comes normalized from $0$ to $1$ and contains masking, that identifies when the captured data was unreliable, or the user was not wearing the device, and so it can be discarded. Regarding the data set organization, it is divided into deep features, in which the features are presented as close to raw as possible, while still being normalized, and handcrafted features which were calculated from not normalized features, but are presented normalized as well.

\subsubsection*{Data Filtering} 

Regarding filtering the data to achieve more representative results, we opt to filter out entries presenting more than half the data as unreliable (in one minute). Not removing these points could impact our output, given that each valid point would have twice the impact on the label result. Another reason for this choice are the experimental results, given that ratio presents the best results as shown in TABLE \ref{tab:trainingDatasetComparison}.

\begin{table}[]
    \centering
    \caption{Accuracy and F1-score metrics in ratio of non-zero values on the training data set.}
    \label{tab:trainingDatasetComparison}
    \begin{tabular}{|c|c|c|}
        \hline
        \rowcolor[HTML]{C0C0C0} 
        Non-zero ratio & {\cellcolor[HTML]{C0C0C0}Accuracy} & {\cellcolor[HTML]{C0C0C0}F1 Score} \\ \hline        
        0.3 & 0.54 & 0.58 \\
        0.4  & 0.54 & 0.58 \\
       \textbf{0.5}  & \textbf{0.56}  & \textbf{0.60} \\
        0.6  & 0.53  & 0.57 \\
        0.7 & 0.53 & 0.56 \\ \hline
    \end{tabular}
\end{table}


\subsubsection*{Testing}
Testing is also an important part of dealing with the data set, since it is the way we validate or discard the hypothesis. For these tests we searched and found k-fold split and the k-fold stratified to be a good compromise between good output and low complexity.

We choose k-fold, since k-fold stratified removes entries to balance classes, and in the case of stress detecting, the order by which the entries are removed is important, since a timeline exists. Our solution is to balance the data set ourselves, while using k-fold. We are sticking with $3$ splits, to try and avoid overfitting while keeping the relation between train set and test set sizes near the real size relation between the train data set ($2060$ entries) and the test data set ($960$ entries).

\subsection*{Features}

The supplied data set contains features extrapolated from the original data set via ML and reassembling, as well as features based on HR, GSR and ST that were extracted from the data and presented as byproducts of the original data. The handcrafted features contain some lost granularity, but the deep features are normalized, and so it would be impossible to recalculate new features from them. With that in mind, our choice is to use only the handcrafted features. 

\subsubsection*{Feature Selection}

From the provided $20$ features on the handcrafted part of the data set, we have tried and tested some, and ended up processing them to create our own features. Here we present $16$ features, some with a scientific paper support, which we cite, and some of them with an empirical evidence basis. Below we present our selected features, with citations in the case they are inspired by another paper.

\begin{itemize}
    \item Heart Rate
    \begin{itemize}
        \item Mean HR \cite{gjoreski2017monitoringstress},
        \item HR standard deviation \cite{gjoreski2017monitoringstress},
        \item HR quartile deviation (percent. 75 - percent. 25) \cite{gjoreski2017monitoringstress},
        \item HRV standard deviation variability \cite{can2019continuousstress},
        \item HRV mean standard deviation\cite{can2019continuousstress},
        \item HRV mean of root mean square of R-R differences,
        \item Percentile 90 of low frequency signal,
        \item Percentile 90 of low and high frequency ratio,
        \item Mean of low and high frequency ratio.
    \end{itemize}
    \item Galvanic Skin Response
    \begin{itemize}
        \item Mean GSR \cite{can2019continuousstress},
        \item GSR quartile deviation (percent. 75 - percent. 25) \cite{can2019continuousstress}.
    \end{itemize}
    \item Skin Temperature
    \begin{itemize}
        \item Mean ST \cite{gjoreski2017monitoringstress},
        \item Mean ST Variability,
        \item Max ST slope value,
        \item ST Mean Slope \cite{gjoreski2017monitoringstress},
        \item Percentile $90$ of ST slope.
    \end{itemize}
\end{itemize}

In the case of heart activity, the unreferenced features, are added due to the fact that low and high frequencies, as well as the ratio between them are good measures of stress related activity \cite{can2019continuousstress}. The non cited data on ST, is used because increases and decreases in ST values can be indicators of stress, and the rate of increase is the biggest indicator, that is why both variability of the mean and 3 slope values are present.

All of these and more features are passed through a ridge correlation between each feature and the label, and none of them had smaller absolute correlations than $20\%$ or bigger than $200\%$, to keep balance. The data with low correlations are kept because they are often referred in literature as good indicators, and since we are still applying a feature extractor tool, not much harm can be done.

\subsubsection*{Feature Extraction}

Since the provided data set contains a minute of data per each label, features must be downsampled, but by doing that we would be losing granularity. In response, our group opted to use research used features while down sampling said data. To try and have the most important features, we tried two feature extractors, that are used with the classifier as algorithm, since we assume the same algorithm is the best feature extractor for itself.

The two tried and tested feature extractors are the recursive feature elimination with cross-validation (RFECV) \cite{iranfar2021relearn}, and the sequential feature selector, which are both provided by the SKLearn framework. They both have been positively used in the literature, and are currently regarded as trustworthy. The TABLE \ref{tab:classifierResultsTable} presents their results using Linear-SVC (C-Support Vector Classification) as the classifier, and being tested with the k-fold split. We use $5$ features minimum, as we think that is the best relation between features and the size of the data set.

\begin{figure}[thpb]
    \centering
    \framebox{\parbox{3in}{\includegraphics[width=3in]{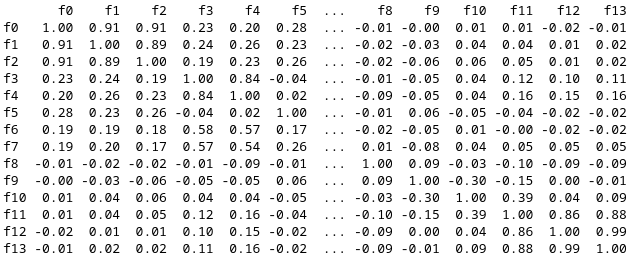}}}
    
    \caption{Screenshot of the RFECV feature correlation results.}
    \label{fig:RFECV}
 \end{figure}

 \begin{figure}[t]
    \centering
    \framebox{\parbox{3in}{\includegraphics[width=3in]{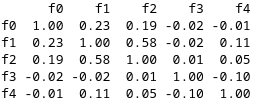}}}
    \caption{Screenshot of the Sequential Feature Selector results.}
    \label{fig:SFS}
 \end{figure}

RFECV gives, on average, more features according to the Figs. \ref{fig:RFECV} and \ref{fig:SFS} with higher correlations between them; we believe that the data set is not big enough for so many features. On the other hand, after testing both options on the test page, the results from the RCEV are $0.59$ accuracy with $0.51$ F1-score against the results from the sequential feature selector which are around $0.62$ accuracy with $0.54$ F1-score, making the sequential feature selector a more suitable choice.

\subsection*{Machine Learning Classification}

Physiological data varies from individual to individual, and while classifying data on a per subject basis can give the ML a personal approach, our data does not have personal identifiers. With that in mind, the error rate is going to be higher, since the values that identify stress in a person are not exactly the same that identify stress on a group.

Furthermore, we approach this challenge using ML algorithms, such as K-NN with multiple neighborhoods (that is 5-NN, 7-NN and 9-NN), SVM, NB, Random Forest and Decision Trees, since those have been fairly covered in the literature, as stated before. K-NN is a method that uses k-nearest data-points and does a majority vote to predict the result, and k is used to identify the number of data-points. SVM finds hyper-planes to divide data-points into different classes \cite{Han2020objectivestress}. We use the SKlearn implementation of SVM, and mostly Linear-SVC. NB classifier predicts the result based on the probabilities of each feature’s probabilistic knowledge, and Random Forest and Decision Trees work by iterating trees of questions and ending with a conclusion in the end.

The model testing evaluation is done with the training data set, as we opt to keep that variable constant. This choice was made in an effort to reduce the complexity of the system, since not having labels for the test data set proved to be a challenge, since its results do not completely correlate with our train data set results.

\section{Results}\label{sec:results}

In this paper, the features without context are used to classify stress. While we have some limitations regarding the data set, we designed a machine learning flow to receive the data and output a label a list of labels classifying stress with a binary classification. Our final algorithm is presented in Algorithm \ref{algo:pseudo}.

\begin{algorithm}
\caption{Pseudo-code of the workflow}
\label{algo:pseudo}
filterDatasetZeros(p=0.5,trainDataset)\\
discretizeDataset(trainDataset)\\
balanceDataset(trainDataset)\\
clf = Classifier\\
selector = SequentialFeatureSelector\\
pickedFeatures = selector.selectFeatures(clf,trainDataset)\\
clf.fit(pickedFeatures)\\
clf.predict(testDataset)\\
output(answer.txt)
\end{algorithm}

Given that, we will now change the classifier for each of the suggested classifiers and access its viability for the challenge. Our findings are presented in TABLE~\ref{tab:trainingDatasetComparison}. Regarding the testing, we perform it on the train data set using k-fold, as well as tests on the challenge submission page.

\begin{table}[]
    \centering
    \caption{Results from different classifiers, using the train and test data set.}
    \label{tab:classifierResultsTable}
    \begin{tabular}{|c||c|c|c|c|}
    \hline
    \rowcolor[HTML]{C0C0C0} 
    {\cellcolor[HTML]{C0C0C0}Methods} & 
    {\cellcolor[HTML]{C0C0C0}K-fold Acc} & 
    {\cellcolor[HTML]{C0C0C0}K-fold F1} & 
    {\cellcolor[HTML]{C0C0C0}Test Acc} & 
    {\cellcolor[HTML]{C0C0C0}Test F1} \\ 
    \cline{1-5} 
    Linear-SVC  & 0.57 & 0.56   & \textbf{0.64}  & 0.54  \\
    Random Forest & 0.47  & \textbf{0.58}  & 0.51   & \textbf{0.58}  \\
    Decision Tree & 0.51 & 0.55   & 0.49  & 0.50 \\
    5-NN  & 0.54  & \textbf{0.58}  & 0.48  & 0.51 \\
    7-NN & 0.53   & \textbf{0.58}  & 0.52   & 0.54 \\
    9-NN & 0.52 & 0.57  & 0.53 & 0.55 \\
    NB  & \textbf{0.58}  & 0.57  & 0.54   & 0.41 \\  \hline
    \end{tabular}
\end{table}

Given this, our pick is using the Linear-SVC, since it presents the best results. The Random Forest is also used initially with a varying degree of success, even allowing us to achieve an accuracy of $0.62$ with specialized parameters. But after tuning, the Linear-SVC provides the best consistent results. While the score, is not high enough to be used in a real setting, it proves promising, given the limitations.

The code is fully available at \url{https://github.com/matpato/CfP-Workshop-and-Challenge-Wellbeing.git}, a sample of the data set is available too.

\subsection{Discussion}
Given the complexity of a stress monitoring solution, better can always be achieved, even though our result is notable among our peers, some things could have been better. The data set being open, and having a test page was a good way of avoiding data overfit, but regarding the data set, some things could be improved.

The data set not having real data was a slight inconvenient, since a lot of nuance was lost, but that did not make it impossible to achieve good predictions. On the other side, the data being normalized is a problem, since we could not extract features from the data, and features that can be interesting could not be used. Examples of this are, inter beat interval (IBI), and its variance on a different time window, the low and high frequencies at full granularity and even the zero of the temperature slope. Those values could have made the scores better, and are commonly used in research, yet we can not calculate them with full precision.

Regarding our work, the usage of SKLearn alone is a problem, since it restricts our
access to machine learning tools. It is possible that deep neural networks can provide better results, but since the time was little to learn a new framework, the group made the choice to play it safe on the framework. More feature tuning and classifier tuning can also have be used to improve results, but this suggestion had issues with computational power when it was tried.

\section{Conclusion}\label{sec:conclusion}
In this paper we propose a stress monitoring system to be applied on the Anxolotl project. While we can adapt the algorithm to different data, namely from the wearable, new data sets need to be tested, to assert its viability. Our results show a $64\%$ accuracy score, which is not high for real life application, but that can be a result of the data set. More research is needed on that regard. While this result is not the best, we are confident that this model has potential to achieve viability on real world classification after improvements.




\bibliographystyle{IEEEtran}
\bibliography{bibliography}

\end{document}